# Optimizing the lattice design for a diffraction-limited storage ring with a rational combination of particle swarm and genetic algorithms

Yi Jiao[†], Gang Xu

Key Laboratory of Particle Acceleration Physics and Technology, Institute of High Energy Physics,
Chinese Academy of Sciences, Beijing 100049, China

**Abstract**: In the design of a diffraction-limited storage ring (DLSR) consisting of compact multi-bend achromats (MBAs), it is challenging to simultaneously achieve an ultralow emittance and a satisfactory nonlinear performance, due to extremely large nonlinearities and limited tuning ranges of the element parameters. Nevertheless, taking the High Energy Photon Source (HEPS) as an example, we demonstrate that the potential of a DLSR design can be explored with a successive and iterative implementation of the particle swarm optimization (PSO) and multi-objective genetic algorithm (MOGA). It turns out that with a hybrid MBA lattice, it is feasible for the HEPS to attain a natural emittance of about 50 pm.rad, and meanwhile, realize a sufficient ring acceptance for beam accumulation with an on-axis longitudinal injection scheme. Particularly, this study indicates that a rational combination of the PSO and MOGA is more effective than either of them alone in approaching the true global optima, for an explorative multi-objective problem with many optimizing variables and local optima.

## 1. Introduction

In the past few decades, the third generation light source (TGLS), based on electron storage ring with a natural emittance of a few nm.rad, has become one of the most widely used platforms providing high quality photon beam for fundamental researches in physics, chemistry, materials science, biology and medicine [1]. Nevertheless, people never stop in pursuing better sources. Early in 1990s, scientists proposed [2] to use multi-bend achromat (MBA) lattices to reduce the natural emittance by at least one order of magnitude to approach the diffraction limit for photons in the energy range of interest (especially in the X-ray range) for user community, so as to push beyond the brightness and coherence available in TGLSs. This new-generation light sources, usually called diffraction-limited storage rings (DLSRs) [3], have only recently become practical and cost effective, with the development of small-aperture magnet and vacuum systems [4-5] and progress in beam dynamics issues. Due to the predicted superior performance of DLSR over TGLS, many laboratories are now constructing storage ring light sources with natural emittance of a few hundred pm.rad (e.g., MAX-IV [6] and Sirius [7]), seriously considering to upgrade existing machines to DLSRs or to build new green-field MBA light sources.

A standard MBA typically has several identical unit cells in the middle and two matching cells on both sides, providing a dispersion-free drift space of a few meters to accommodate the insertion devices (IDs) dedicated to emission of high-flux photon beam. To attain an ultralow emittance with a compact layout, it is common [8] to use combined-function dipoles and strong quadrupoles to achieve small optical parameters that are close to the theoretical minimum emittance (TME [9]) conditions in the dipoles. In some standard-MBA designs [e.g., 10, 11] the emittance is further reduced with damping wigglers. In order to correct the large natural chromaticities arising from the strong focusing, sextupoles are usually located in every unit cell. Even so, it was found [12] that the sextupole strengths scale approximately inversely linearly with the natural emittance. Experience [13] indicated that if reducing the natural emittance to a few tens of pm.rad with standard MBAs, unpractically high sextupole gradients or very thick sextupoles (e.g., thicker than quadrupoles) will be needed, leading to an undesirable DLSR design.

To make a more effective chromatic correction, a particular MBA concept, deemed 'hybrid MBA' [14] was proposed. For example, in a hybrid 7BA (c.f. Fig. 1), the four outer dipoles are used to create two dispersion bumps with much larger dispersions than available in a standard MBA, and all the chromatic sextupoles are placed in the dispersion bumps. In this way, the sextupole strengths can be reduced to an achievable level with conventional magnet technology. Whereas the cost of doing so is that the optical functions are not optimized for emittance minimization. To compensate this effect, even stronger focusing than in a standard MBA is adopted neighboring the inner three dipoles and longitudinal gradients are introduced to the outer dipoles to achieve an ultralow emittance. Moreover, the optics is matched to form a $-I$ transportation between each pair of sextupoles to cancel most of the nonlinearities induced by the sextupoles. Due to these advantages, the hybrid MBA lattice has been adopted in many projects, e.g., ESRF-II [14], Sirius [7] and APS-U [15].

A kilometre-scale storage ring light source with a beam energy of 5 to 6 GeV, named the High Energy Photon Source (HEPS), is to be built in Beijing, China. The lattice design has been continuously evolved. Recently a hybrid 7BA design with a natural emittance of 60.1 pm·rad at 6 GeV was developed [16]. This design (denoted as 'mode I' hereafter) consists of 48 identical hybrid 7BAs and has a circumference of 1296.6 m. Each 7BA is of about 27 m, with a 6-m ID section. The layout is very compact. Several gaps between dipoles and adjacent quadrupoles (D3, D4 and D8 in Fig. 1) are of only 7.5 cm. And, to reserve as much space as possible for diagnostics and other equipment, in each 7BA only six sextupoles and two octupoles are located in the two dispersion bumps for chromatic correction and nonlinear optimization. The layout and optical functions of a 7BA are shown in Fig. 1, and the main parameters of the ring are listed in Table 1.

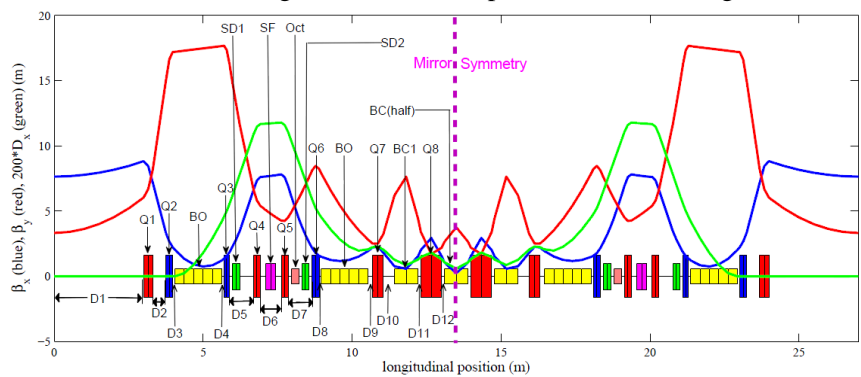

Fig. 1. The optical functions and the layout of a hybrid 7BA of the HEPS 'mode I' design. Due to the mirror symmetry, only the elements on the left sides of the 7BA are marked with names.

Table 1. Main parameters of the original and optimized HEPS designs

|  | **Mode I** | **Mode II** | **Mode III** | **Mode IV** | **Mode V** |
|---|---|---|---|---|---|
| Working point (H/V) | 113.20/41.28 | 116.16/41.12 | 111.28/41.11 | 112.13/41.18 | 112.23/39.14 |
| Natural chromaticity (H/V) | −149/−128 | −214/−133 | -134/-136 | −140/−137 | −140/−126 |
| Beta functions in ID section (H/V)/m | 7.6/3.3 | 9/3.2 | 7/3.1 | 8.0/3.1 | 7.9/4.2 |
| Natural emittance $\varepsilon_0$/(pm·rad) | 60.1 | 59.4 | 58.9 | 55.8 | 52.1 |
| Lengths of (SD/SF/OF)/m | 0.25/0.34/0.26 | 0.25/0.34/0.26 | 0.22/0.25/0.2 | 0.22/0.25/0.2 | 0.22/0.25/0.2 |
| $K_s/K_{oct}$ w/ 0.2-m multipoles/ ($m^{-3}/m^{-4}$) | 294/3.3 ×$10^4$ | 255/1.5 ×$10^4$ | 273/1.3 ×$10^4$ | 273/1.4 ×$10^4$ | 284/1.3 ×$10^4$ |
| Momentum acceptance | 2.4% | 3% | 3.6% | 3.6% | 3.5% |
| Horizontal and vertical DA size/mm | 2.5/2.2 | 2.5/3.5 | 3.0/4.4 | 3.5/4.5 | 3.0/4.8 |

To evaluate the nonlinear performance of a 'realistic' machine, it is necessary to consider various errors (such as misalignments, rotations, strength errors, etc.) in the lattice evaluation. One needs to generate a large number of random seeds for errors, add them to the bare lattice, simulate the lattice calibration procedure, and finally calculate the ring acceptance, i.e., the dynamic aperture (DA) and momentum acceptance (MA). Such a process, however, is always complex and time-consuming. Instead, in HEPS design we calculated the 'effective' DA and 'effective' MA of the bare lattice, and used them as indicators of the nonlinear performance. Within the effective DA or MA, it is required not only the motion remains stable after tracking of a few thousand turns, but also the tune footprint is bounded by the integer and half integer resonances nearest to the working point (c.f. Fig. 2). The thought behind this definition is that the integer and half integer resonances can be excited by linear field imperfections, being fatal to beam dynamics. Although some recent experimental studies in TGLSs [e.g., 17] showed half integer resonances can be safely approached or even crossed with the state-of-art optics correction technique, detailed simulation studies for the HEPS design indicated that the half integer resonances can cause particle loss even with a tiny focusing error. Thus, it is believed that the effective DA and MA provide a reasonable measure of the ring acceptance of a DLSR in presence of machine imperfections.

For the 'mode I' design, the multipoles were grouped in four families (SD1, SD2, SF and Oct). Two sextupole families were for chromatic correction, and only two free knobs were left for the DA and MA optimization. This enabled us to perform a grid scan of the multipole strengths in a reasonable computing time, based on numerical tracking with the AT [18] program and frequency map analysis [19]. Unfortunately, it was found difficult to simultaneously optimize the effective DA and MA. The compromise solution predicts an effective DA of 2.5 (or 2.2) mm in the *x* (or *y*) plane and an effective MA of 2.4%, with the results shown in Fig. 2 and 3.

---

† jiaoyi@ihep.ac.cn

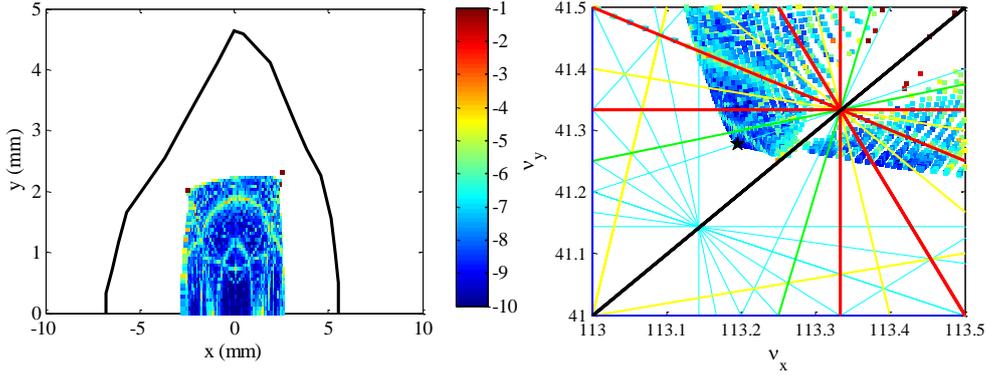

Fig. 2. The effective DA and the corresponding frequency map for the HEPS 'mode I' design. The colors, from blue to red, represent the stabilities of the particle motion, from stable to unstable. The DA, with all surviving particles after tracking over 1000 turns, is also plotted (black curve) for comparison. For this and the following results, the bare lattice was always used in the numerical tracking.

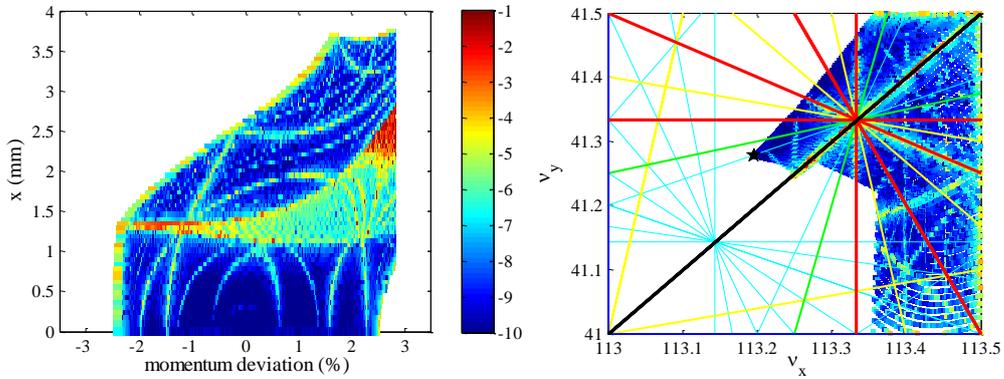

Fig. 3. The effective MA and the corresponding frequency map for the HEPS 'mode I' design. The colors, from blue to red, represent the stabilities of the particle motion, from stable to unstable.

To resolve the injection difficulty caused by the small DA, we proposed [20] a novel on-axis longitudinal injection scheme enabled by phase manipulation of a double-frequency RF system. Compared with that with a single-frequency RF system [21], this scheme can greatly reduce the requirement of the MA to about 3%. We believe that it is possible, although challenging, to reach such a target on MA. Actually, by means of sextupole strength minimization and tune space scan for the same layout as the 'mode I' design, we attained a better design [22] (denoted as 'mode II' hereafter) with a similar emittance, 59.4 pm.rad, but larger effective MA (~3%) and DA (~2.5 mm in $x$ and 3.5 mm in $y$ plane). The main parameters of the 'mode II' design are also listed in Table 1.

However, before taking the 'mode II' lattice as the 'final design', it is necessary and important to globally scan all the tunable element parameters (while keeping the circumference basically unchanged, i.e., varied in +/− 1 m) to explore the ultimate performance of such a hybrid 7BA lattice. The performance parameters include the achievable minimum natural emittance, and the maximum ring acceptance at a specific natural emittance. It is also interesting to investigate the dependence of the nonlinear performance on various linear optical parameters.

For a hybrid 7BA, there are more than 20 tunable element parameters. A global grid scan may take too long a time to exhaust all the possibilities. In contrast, a more efficient way is to use stochastic optimization algorithms, e.g., the multi-objective genetic algorithm (MOGA) and multi-objective particle swarm optimization (MOPSO). The MOGA methods mimic the process of natural selection and evolution of species, and have been widely applied to many accelerator optimization problems [23-27]. While MOPSO emulates the self-organizing behavior of social animal living in group, and has been recently used to optimize the linac operation and nonlinear dynamics of storage rings [28-30].

It has been demonstrated that both algorithms are powerful and effective in solving the problems with piecewise continuous and highly nonlinear objectives and many local optima. Nevertheless, a recent study [30] showed that MOPSO converges faster than MOGA, and is not as dependent on the distribution of initial population as MOGA. To test this, we compared the performance of these two algorithms by applying them to a problem with a known answer. The results are presented in Sec. 2. It was found that each algorithm has its own unique advantage, and implementing them in a successive and iterative way will be more effective than using either of them alone in approaching the true global optima for an explorative multi-objective problem. As will be shown in Sec. 3, with such a combination of MOGA and MOPSO, we were able to find solutions showing optimal trade-offs between the natural emittance and ring acceptance for the HEPS hybrid 7BA design. Conclusions will be given in Sec. 4.

## 2. Optimization of the natural emittance and chromatic sextupole strengths

### 2.1 Optimization for the case with a fixed ID section length

Experience [22] indicated that sextupole strength minimization followed by tune space scan is an effective way to improve the nonlinear performance of a DLSR. Thus, we first looked at the trade-offs between the natural emittance and the sextupole strengths required for chromaticity correction, for the case with a fixed ID section length, $L_{\mathrm{ID}} \equiv 6$ m. Since the evaluation limits itself in linear optics regime and is very fast (less than 1 s per evaluation), we could obtain optimal solutions in a much shorter time than that takes to directly optimize the effective DA and MA.

One of the MOGA methods, non-dominated sorting genetic algorithm II (NSGA-II [31]), was used in the optimization. Totally 26 element parameters were used as optimizing variables, and varied within specific ranges that are determined by practical or optical constraints (see Table 2 for details). Two objective functions, weighted natural emittance $\varepsilon_0$ and weighted chromatic sextupole strengths, were defined. To compare the sextupole strengths between different solutions, the sextupoles were grouped in just two families (SD, SF) with identical lengths of 0.2 m, such that for specific corrected chromaticities ([0.5, 0.5] in this study) there is a unique solution of the sextupole strengths ($K_{sd}$, $K_{sf}$), which were then represented with a nominal strength,

$$K_s = \sqrt{(K_{sf}^2 + K_{sd}^2)/2}. \tag{1}$$

To ensure enough diversity in the initial population, we first randomly generated lots of possible combinations of optimizing variables with large enough fluctuations around those of the 'mode I' and 'mode II' designs, from which we selected 6000 solutions with stable optics and used them as the initial population. This took a long, but still acceptable, computing time.

Table 2. Optimizing variables and scanning range in the optimization

| Variables | Scanning range |
|---|---|
| Lengths of the drifts | [0.1, 1.6] m |
| Gradients of (Q1 to Q6) | [-2.6, 2.6] m$^{-2}$ |
| Gradients of (Q7 and Q8) | [-4, 4] m$^{-2}$ |
| Gradients of dipoles (BC1 and BC2) | [-2.4, 2.4] m$^{-2}$ |
| Length of inner dipoles | [0.6, 1.0] m |
| Length of outer dipoles | [1.2, 1.8] m |
| Dipole angles | [0.1, 2] degree |

For each ensemble of variables, before evaluating the $\varepsilon_0$ and $K_s$, several quadrupoles were tuned to match (if feasible) the achromatic condition (with $K_{Q3}$ and $K_{Q4}$) and the $-I$ transportation between each pair of sextupoles (with $K_{Q5}$, $K_{Q6}$ and $K_{Q7}$). Moreover, to ensure that the obtained solutions have desirable optics, as many as possible constraints were considered:

(1) a reasonable maximum value of beta function along the ring, $\max(\beta_{x,y}) \leq 30$ m;
(2) reasonably low beta functions in ID section for high brightness, 1.5 m $\leq \beta_y <$ 4 m and 1.5 m $\leq \beta_x <$ 15 m;
(3) stability of the optics, $\text{Tr}(M_{x,y}) < 2$, with $M_{x,y}$ being the transfer matrix of the ring in the $x$ or $y$ plane;
(4) fractional tunes in (0, 0.5), which is favorable against the resistive wall instability;
(5) reasonable natural chromaticities, $|\xi_{x,y}| \leq 5.5$ in one 7BA;
(6) all drifts between adjacent magnets longer than 0.1 m;
(7) one of the drifts (D10, D11 and D12) longer than 0.35 m to accommodate a three-pole wiggler, which is to be used as a hard X-ray source;
(8) reasonably low energy loss in each turn due to synchrotron radiation ($U_0 \leq 2.2$ MeV).

The degree of the violation of each constraint was measured with a weight factors. If a specific constraint is satisfied, the corresponding weight factor will be one; otherwise the factor will be assigned a value above 1. The more violated the constraint is, the larger the factor will be. These factors were then multiplied by the calculated $\varepsilon_0$ and $K_s$ to get the values of the two objective functions. In this way, even with similar or the same $\varepsilon_{x0}$ and $K_s$, the desirable solutions (meet all constraints) will have smaller objective functions than those that violate certain constraints, and will be assigned a higher rank with the non-dominated sorting, and have higher priorities for survival and reproduction in the evolution chain.

It is worth mentioning that in the optimization only the quadrupole gradients, rather than both the gradients and lengths, were used as optimizing variables. This is based on the consideration that for a specific change in the transverse focusing, there will be a myriad of possible combinations of the quadrupole lengths and gradients, whereas we are just interested in the solutions which have shortest possible quadrupoles. Therefore, if both quadrupole lengths and gradients are varied in the optimization, it probably requires an additional sorting of the solutions, and needs to evolve a larger population over more generations.

Instead, we optimized the quadrupole strengths with an iteration of the MOGA algorithm. In the first MOGA evolution, the quadrupoles had the same lengths as in the 'mode I' design, while their gradients were varied in larger ranges than available. According to the covering range of gradients of the final population, we adjusted the quadrupole lengths in such a way that all the gradients are below but close to their upper limits. And then, the final population of the first MOGA (with small modifications on gradients, if necessary) was used as the initial population of a new MOGA evolution, where the quadrupoles had modified values and their gradients were varied within the available ranges. The population evolved over 1000 generations, and as shown in Fig. 4, the population was already very close to the final population at generation 600. The figure also shows that for the HEPS hybrid 7BA design, it is feasible to reduce the natural emittance to about 43.5 pm.rad, or to decrease the nominal sextupole strength $K_s$ to about 180 m$^{-3}$ at $\varepsilon_0 = 60$ pm.rad, which is much smaller than those in the 'mode I' and 'mode II' designs (294 and 255 m$^{-3}$ with 0.2-m sextupoles).

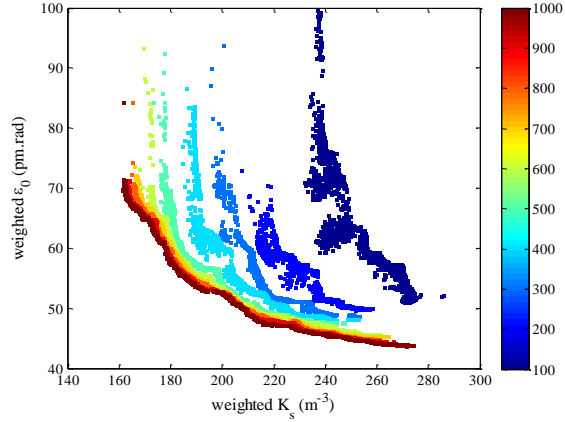

Fig. 4. Objective functions of the population at every hundred generation (in different color) with MOGA.

Among the solutions of the 600th to 1000th generation, we kept only the desirable solutions with $\varepsilon_0$ below 65 pm.rad, and then re-evaluated them to get other optical parameters, including the beta functions at ID section, dispersion at dispersion bump ($D_x$), tunes and natural chromaticities. The distributions of these solutions in different sub-parameter planes are shown in Fig. 5.

In these solutions, the nominal sextupole strength $K_s$ decreases monotonously with increasing $\varepsilon_0$ and increasing $D_x$. Besides, many optical parameters rely heavily on the integer tunes, especially the horizontal one. For instances, the covering ranges of the beta functions at ID section are different in different integer tune regions. And, a larger horizontal integer tune corresponds to larger natural chromaticities, larger dispersion at dispersion bump, weaker sextupoles and a larger emittance. Although both the dispersion at dispersion bump and natural chromaticities contribute to the nominal sextupole strength, which, apparently, is more related to the former factor. Further study revealed that to achieve a larger $D_x$, one needs to match the optics in such a way that the optical functions in the four outer dipoles are more deviated from the TME conditions, leading to an increase in emittance. To keep the emittance in an ultralow level, stronger focusing (a larger horizontal tune) is needed to squeeze the optical functions in the inner three dipoles and get them closer to the TME conditions. It appears that a larger horizontal integer tune helps to reach a balance between ultralow emittance and weakest possible sextupoles.

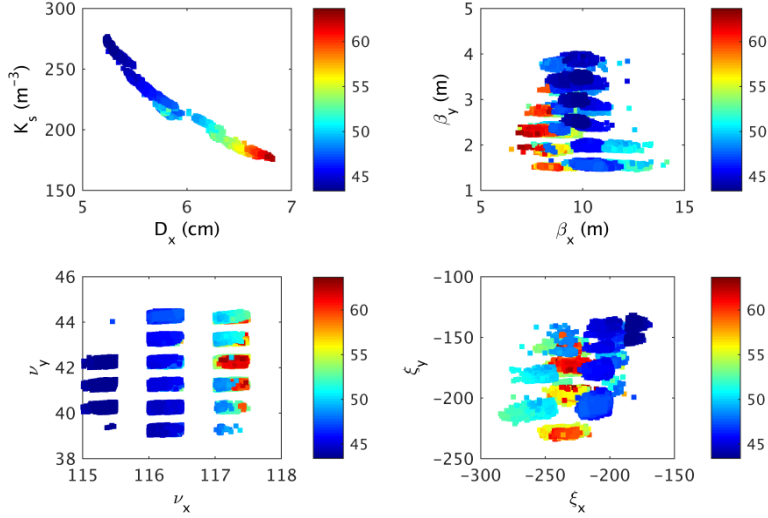

Fig. 5. Selected MOGA solutions projected onto the $K_s$-$D_x$, $\beta_x$-$\beta_y$, $\nu_x$-$\nu_y$ and $\xi_x$-$\xi_y$ planes. The colors represent the values of the natural emittance (in unit of pm.rad).

## 2.2 Optimization for the case with a variable ID section length

In the above optimization, the ID section length was fixed to 6 m. One can consider that if with a shorter $L_{ID}$, the variables for the position and length of magnets will have larger adjustment space, and it would be feasible to achieve designs with better performance.

To explore the potential of the design with a shorter ID section length, in this optimization the $L_{ID}$ was also used as an optimizing variable, and varied in the range of [5, 7] m. The final population of MOGA obtained in Sec. 2.1, with small modifications, was used as the initial population of this optimization. The modifications included generating random values drawn from a normal distribution with an average of 6 m for the $L_{ID}$, and accordingly adjusting the length of the drift D6 to keep the circumference unchanged. The standard deviation of the random seeds for $L_{ID}$ was, however, set to a small value (0.1 m), to ensure that most of the individuals in the initial population have stable optics.

For comparison, the same initial population was evolved over 800 generations with MOGA and MOPSO, respectively. The parameter settings of these two algorithms are the same as described in Ref. [30]. It was found that the solutions with $L_{ID}$ larger than 6 m were phased out in the evolution with both algorithms. In addition, as shown in Fig. 6, most of the solutions at the last generation of MOGA and MOPSO have better performance (e.g., with smaller $K_s$ at a specific $\varepsilon_0$) than those optimized for $L_{ID} \equiv 6$ m.

On the other hand, the difference in the performance of these two algorithms is also obvious. For MOGA, the $L_{ID}$ values of the final population do not exceed the $L_{ID}$ covering range of the initial population, with a minimum of about 5.75 m. While for MOPSO, a majority of solutions have $L_{ID}$ values close to 5 m, and in particular, predict smaller $K_s$ than those obtained with MOGA. It was noticed that this difference had been well explained in Ref. [30]. In MOPSO, each surviving individual adjusts its moving pace and direction in parameter space at every iterative step, according to its own historical experience and relative position within the population. The new solutions are not generated from the existing good solutions, as what is done in MOGA. Thus, MOPSO intrinsically allows more diversity than MOGA, and does not need a diverse seeding in the initial population.

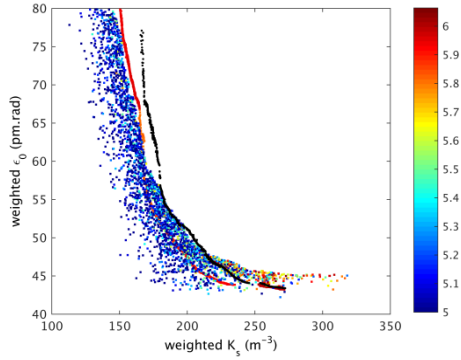

Fig. 6. MOGA solutions for fixed ID section length, $L_{ID} \equiv 6$ m (black curve), and the solutions with MOPSO and MOGA for variable $L_{ID}$, with the colors representing the $L_{ID}$ values (in unit of m).

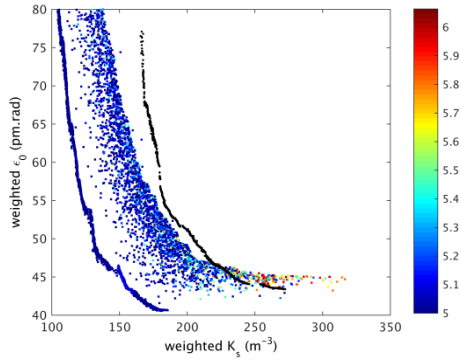

Fig. 7. MOGA solutions for fixed ID section length, $L_{ID} \equiv 6$ m (black curve), and the solutions for variable $L_{ID}$ after evolution of 500 more generations with MOGA and MOPSO, with the colors representing the $L_{ID}$ values (in unit of m).

Nevertheless, one can see from Fig. 6 that the final solutions of MOPSO distribute rather sparsely in the objective function space. Also, in the low emittance region ($\varepsilon_0 \sim 45$ pm.rad), some solutions still have $L_{ID}$ values close to 6 m, with even larger $K_s$ than those optimized for $L_{ID} \equiv 6$ m. As shown in Fig. 7, the situation does not substantially change even after 500 more generations of evolution with MOPSO. Just for comparison, based on the MOPSO population at generation 800, we also implemented MOGA for 500 generations. One can see from Fig. 7 that the final population of MOGA reached a better convergence to the Pareto optimal front, predicting solutions with all $L_{ID}$ values close to 5 m and with superior performance over those optimized for

$L_{ID} \equiv 6$ m in the whole emittance range of interest.

By optimizing the design first with MOPSO and then with MOGA, we obtained solutions that accord with expectation. It appears that by shortening the ID section from 6 m to about 5 m, the natural emittance can be further reduced to about 40 pm.rad, or the nominal sextupole strength $K_s$ can be further decreased by at least 40% at $\varepsilon_0 = 60$ pm.rad. On the other hand, it is worthy to mention that a shorter $L_{ID}$ implies a reduced space for IDs, which may affect the performance of the light source. Therefore in the following we will only discuss the case with $L_{ID} \equiv 6$ m.

Nevertheless, from the above results one can learn that MOGA depends significantly on the distribution of initial population. If without enough diversity in the initial population, MOGA may converge to local optima rather than the true global optima. Worse still, the MOGA itself cannot give a measure of the diversity of a population. Consequently, if applying MOGA to a typical exploratory multi-objective problem with many optimizing variables and local optima, and without another effective algorithm (e.g., MOPSO in this study) for comparison, one cannot know for sure whether the final solutions reveal optimal trade-offs between the different objectives. In short, to make an effective MOGA optimization, it is critical, and also challenging, to seed the initial population with high enough diversity. Fortunately, as demonstrated above, this difficulty can be overcome with the MOPSO, which has an intrinsic ability of breeding more diversity in the evolution of population. And once the diversity of solutions is ensured, MOGA can reach a better convergence than MOPSO to the true global optima. Therefore, evolving the population with a rational combination of MOPSO and MOGA would be more effective than using either of these two algorithms alone.

## 3. Optimization of the natural emittance and ring acceptance

### 3.1 Nonlinear performance of the solutions with minimized sextupole strengths

In Sec. 2.1, we have obtained solutions showing optimal trade-offs between the natural emittance $\varepsilon_0$ and the chromatic sextupole strengths (represented with a nominal strength $K_s$) for the case with $L_{ID} \equiv 6$ m.

The corresponding nonlinear performance of these solutions was then evaluated. For simplicity, the nonlinear performance was measured with the scaled DA area (in unit of mm$^2$), i.e., the product of the horizontal and vertical effective DA sizes normalized with respect to the square root of the values of beta functions at the start point of DA tracking. Among the solutions with the same or very similar tunes, the one with the lowest $K_s$ was selected. The multipoles were split into four families again, and their strengths were scanned with small step sizes (5 m$^{-3}$ for sextupoles and 100 m$^{-4}$ for octupoles). The available scaled DA area for a specific set of tunes was obtained through numerical tracking for the multipolar set (if exist) that results in an effective MA of not less than 3% and the largest scaled DA area.

To look at the relations between the scaled DA area and other parameters, such as $\varepsilon_0$, $K_s$ and tunes, the solutions were separated into six parts, as shown in Fig. 8(a). In each part, the solutions covered a large range of tune area, and the available maximum scaled DA area was obtained by comparing those with different tunes. The results are shown in Fig. 8(b). It appears feasible to find solutions with scaled DA area larger than that of 'mode II' design (1.63 mm$^2$), and meanwhile, with $\varepsilon_0$ below 60 pm.rad and effective MA equal to or above 3%. Nevertheless, the figure does not show a monotonous variation of the scaled DA area with $\varepsilon_0$ as for the $K_s$. The available scaled DA area also depends on the horizontal integer tunes. This indicates that the sextupole strength minimization followed by tune space scan may be not the best way to find the optimal trade-offs between the natural emittance and ring acceptance. A systematic scan of all the tunable element parameters is necessary. Nevertheless, the obtained solutions provide a good start point for the new optimization, which uses the natural emittance and ring acceptance directly as optimizing objectives.

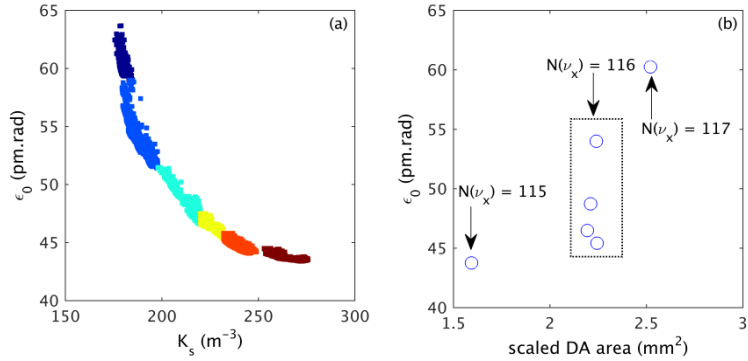

Fig. 8. Solutions projected onto the $K_s$-$\varepsilon_0$ plane, and the available maximum scaled DA area at different emittance range.

## 3.2 Direct optimization of the natural emittance and ring acceptance

In this optimization, the multipole strengths were also used as optimizing variables. The upper limit of the sextupole strength is set to 280 m$^{-3}$, by assuming a larger pole radius for the sextupoles (14 mm) than quadrupoles (12.5 mm), for the sake of extracting photon beam from the upstream IDs. And, the multipoles were split into eight families to provide six free knobs for nonlinear optimization. The nonlinear performance, at this time, was measured with the scaled ring acceptance (in unit of mm$^2$), i.e., the product of the scaled DA area and the effective MA (normalized by 3%). Totally 32 optimizing variables (all tunable element parameters except $L_{ID}$) and two objective functions (weighted natural emittance and scaled ring acceptance) were used.

The individuals of the initial population were selected from the solutions obtained in Sec. 3.1 that promise larger scaled DA area than the 'mode II' design. Although the objective evaluation in this case took a much longer time (~ 60 s per evaluation) than for the evaluation just in linear optics regime, we chose a relatively large population size of 4000 to ensure the comprehensiveness of the solutions.

In addition, to make the obtained solutions have robust nonlinear performance, more constraints were considered. At this time, the effective DA or MA is determined by the amplitude or momentum deviation with tunes first approaching the integer resonances by 0.05 or the half integer resonances by 0.01. It was noticed that the space charge effect can cause a maximum vertical tune shift of about 0.01 for HEPS with a beam current of 200 mA. To avoid being trapping by coupling resonances due to the space charge effect, it is required that the fractional tunes for any momentum deviation should be separated by at least 0.015.

The sextupole lengths were optimized by iterative implementations of MOPSO and MOGA, similar to what was done for optimizing the quadrupole lengths. During the iterations, we also gradually reduced the emittance range of interest (if the calculated natural emittance exceeds the range, the objective functions will multiplied or divided by an addition factor above 1), such that more and more solutions had natural emittance of about 60 pm.rad or even lower. Particularly, it was empirically found essential to evolve the population with MOPSO over enough generations (1000 generations in our study), so as to generate solutions with diverse characteristic parameters. Otherwise, the subsequent MOGA will quickly converge to the local optima, with solutions gathered in a few small distinct regions in the objective function space.

In spite of limited tuning ranges of the optimizing variables and various constraints in the optimization, after several iterations of MOPSO and MOGA, nearly continuously distributed solutions in the objective function space were obtained, showing almost a monotonous variation of the scaled ring acceptance with the natural emittance. The population evolutions at the last iteration of MOPSO and MOGA are shown in Fig. 9. For the final population of MOGA, there is a turning point around $\varepsilon_0 = 50$ pm.rad. The available ring acceptance decreases rapidly with the emittance for $\varepsilon_0$ below 50 pm.rad, while changes with a much smaller slope for $\varepsilon_0$ above 50 pm.rad. This suggests that for the HEPS hybrid 7BA design, it is best to keep the natural emittance above 50 pm.rad to achieve a robust nonlinear performance, i.e., with a high tolerance to small deviations in the linear optical parameters.

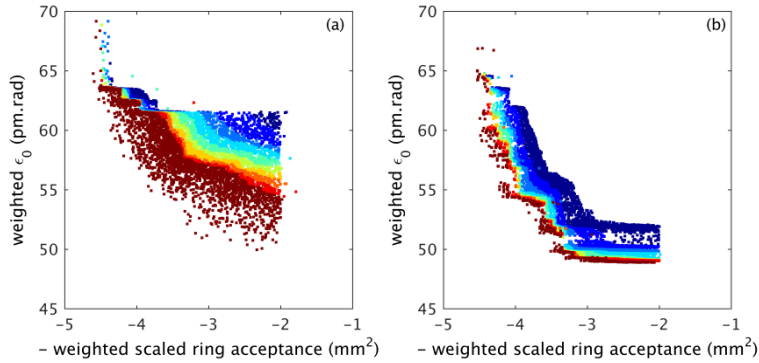

Fig. 9. Solutions of the last iteration of MOPSO (a) and MOGA (b) in the objective function space. The population is plotted at every 100 generation and marked with different colors.

The PSO solutions from generation 500 to 1000 and MOGA solutions from generation 200 to 1000 in Fig. 9 were selected and re-evaluated. From these solutions, only those with scaled ring acceptance above 2 mm$^2$, MA above 3% and $\varepsilon_0$ below 65 pm.rad were kept for post analysis. Fig. 10 shows the distributions of the selected solutions in different sub-parameter planes.

These solutions cover a horizontal integer tune range of [111, 112], which, however, is entirely different from that of the initial population (from 115 to 117). And, these solutions use stronger sextupoles than those in the initial population. Further study showed that the initial solutions with larger horizontal integer tunes were phased out with MOPSO due to their relatively smaller scaled ring acceptance. It appears that weakest possible sextupoles do not lead to a largest possible ring acceptance, and the level of sextupole strengths is neither the decisive nor exclusive factor of the nonlinear performance. In contrast, the level of the transverse focusing (reflected in the horizontal integer tune and natural chromaticities) also has significant impact on the available ring acceptance. In addition, it was demonstrated again that MOPSO is powerful in generating new solutions with different characteristic parameters from the existing ones. One can also see from Fig. 10 that the fractional tunes are dominant factors of the effective MA. By choosing fractional tunes at the bottom left area of the tune space, one can increase the effective MA up to a maximum of about 3.8%.

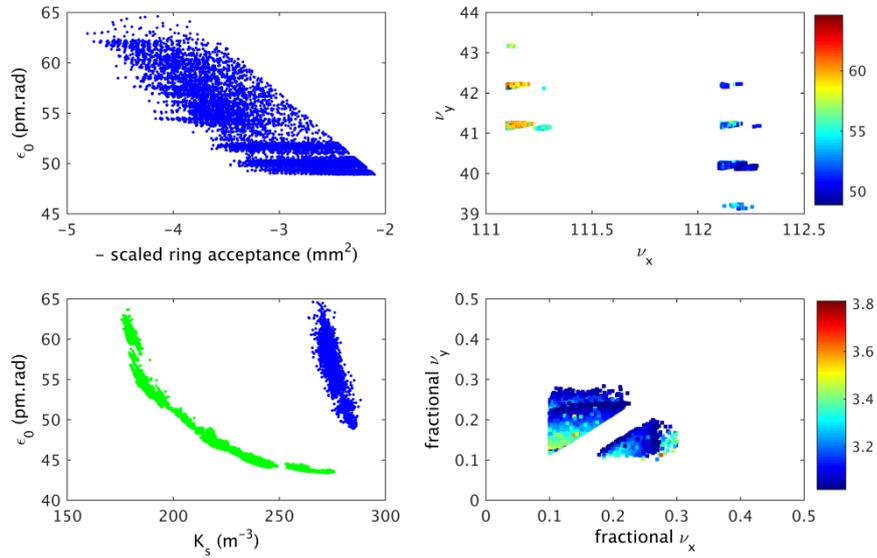

Fig. 10. Selected MOPSO and MOGA solutions projected onto different sub-parameter planes. In the upper right plot the colors represent values of the natural emittance (in unit of pm.rad), in the lower left plot the green dots represent the solutions with minimized sextupole strengths, and in the lower right plot the colors represent values of the effective MA (in unit of %).

As mentioned, for HEPS it was proposed to use on-axis longitudinal injection, which has a stringent demand for MA but a less strict requirement on DA. Thus, the main goal of the nonlinear optimization is to attain as a large effective MA as possible, while keeping the effective DA large enough for on-axis injection (e.g., greater than ten times of the rms transverse beam size at the injection point). To this end, for each specific integer tune region, we did detailed numerical tracking and FMA for the solutions with the largest effective MAs, rather than those with the largest scaled ring acceptances. Several solutions were found and denoted as 'mode III', 'mode IV' and 'mode V'. The main parameters of these designs are also listed in Table 1. Compared to the 'mode I' or 'mode II' design, these designs have lower natural emittance and larger MA, and shorter quadrupoles and multipoles. Also, in these designs the fractional tunes are well separated, and all the drifts are longer than 0.1 m. As a demonstration, the effective DA and MA of the 'mode V' design are shown in Fig. 11 and 12. Note that in Fig. 11 the frequency map gets folded at y ~

2.5 mm. It was reported [32] that the fold corresponds to a singularity in the frequency map, and after the fold, directions of fast escape may appear, causing large diffusion of trajectories. Nevertheless, a vertical acceptance of 2.5 mm is already enough for the on-axis longitudinal injection.

It is worthy to mention that for the solutions with integer tune of (112, 40), the coupling resonance $2\upsilon_x - 2\upsilon_y = 48 \times 3$ is a low order structural resonance and cause a large perturbation to dynamics. Thus, although the solutions in this integer tune region promise both large MA (up to 3.8%) and ultralow emittance (close to 50 pm.rad), which were not chosen as candidate optimal designs for the HEPS.

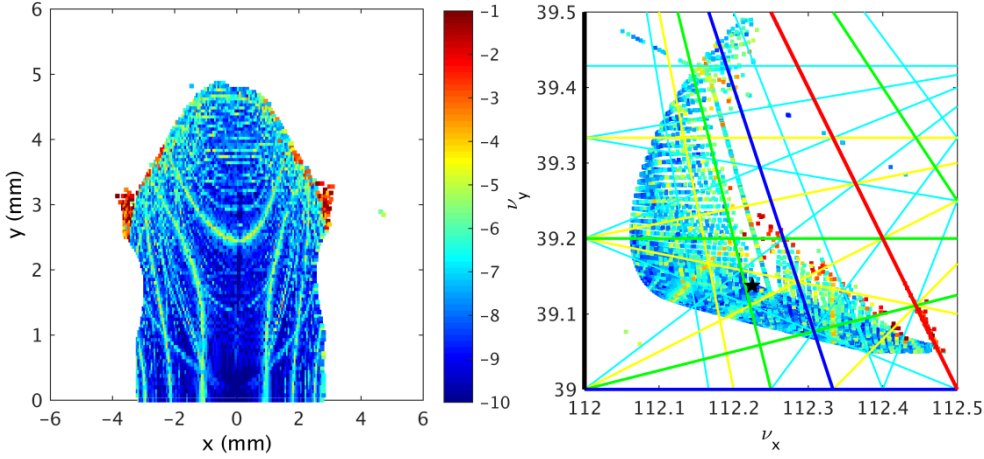

Fig. 11. The effective DA and the corresponding frequency map for the HEPS 'mode V' design. The colors, from blue to red, represent the stabilities of the particle motion, from stable to unstable.

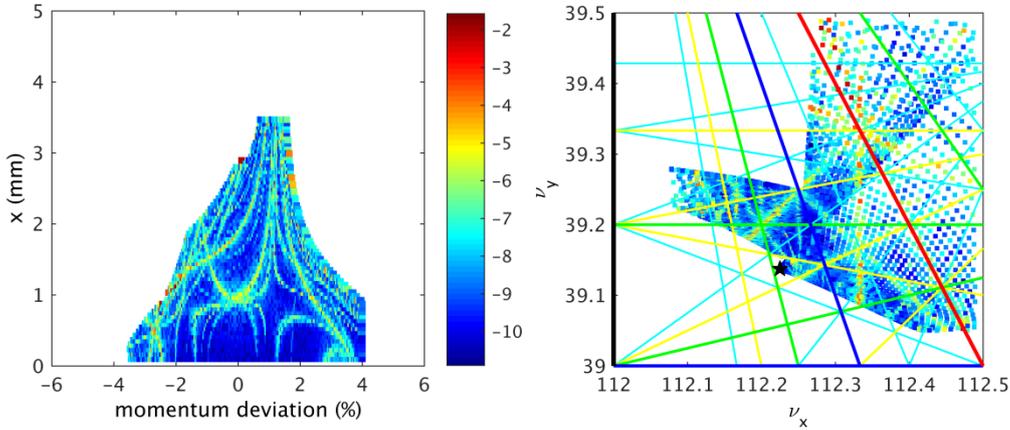

Fig. 12. The effective MA and the corresponding frequency map for the HEPS 'mode V' design. The colors, from blue to red, represent the stabilities of the particle motion, from stable to unstable.

### 4. Conclusion

In this paper, by using a successive and iterative implementation of MOPSO and MOGA, we explored the potential of the hybrid 7BA lattice design for the HPES project from an original design with a natural emittance of 60.1 pm.rad. It appears that with a hybrid 7BA lattice, it is feasible for HEPS to achieve a sufficient ring acceptance for beam accumulation with longitudinal injection, and meanwhile, reduce the natural emittance to about 50 pm.rad.

In the optimization, we also investigated the relations between nonlinear performance and linear optics. We showed that there is neither a decisive nor exclusive factor of the nonlinear performance. In contrast, to simultaneously attain an ultralow emittance and a largest possible ring acceptance, it needs to reach a balance among various factors, especially the sextupole strengths and the integer and fractional tunes.

In short, we showed an effective way how to explore the potential of a MBA lattice from a specific design, while not necessarily requiring a deep understanding of the physics behind the lattice design and the complicated relations between the nonlinear dynamics and linear optics. The key point is to evolve a large enough population with MOPSO and MOGA in a successive and iterative way. As demonstrated, MOPSO has an intrinsic ability of breeding more diversity in the population during evolution. And once with enough diversity in the population, MOGA can reach better convergence than MOPSO. Thus, combining MOPSO and MOGA in the optimization will be more effective and powerful, than using either of the two algorithms, in searching the global optima, regardless of whether or not enough diversity is seeded in the initial population. It is believed that such an optimization procedure can benefit other DLSR designs with the same or similar objective functions (e.g., using DA and MA in presence of magnetic errors instead of the effective DA and MA, or using Touschek lifetime instead of MA), and can be generalized to other explorative multi-objective optimization problems.

**Acknowledgment**
This work was funded by the by the Natural Science Foundation of China (No. 11475202 and 11405187) and Youth Innovation Promotion Association CAS (2015009).

**References**
[1] Zhao Z 2010 *Reviews of Accelerator Science and Technology* **3** 57
[2] Einfeld D, Schaper J and Plesko M 1995 *Proc. PAC'95* TPG08 (*Dallas, USA*)
[3] Hettel R 2014 *J. Synchrotron Rad.* **21** 843-855
[4] Al-Dmour E *et al* 2014 *J. Synchrotron Rad.* **21** 878-883
[5] Johansson M *et al* 2014 *J. Synchrotron Rad.* **21** 884-903
[6] Tavares P F *et al* 2014 *J. Synchrotron Rad.* **21** 862-877
[7] Liu L *et al* 2014 *J. Synchrotron Rad.* **21** 904-911
[8] Jiao Y, Cai Y and Chao A W 2011 *Phys. Rev. ST Accel. Beams* **14** 054002
[9] Teng L C 1984 Fermilab Report No. TM-1269
[10] Cai Y *et al* 2012 *Phys. Rev. ST Accel. Beams* **15** 054002
[11] Jiao Y and Xu G 2013 *Chin. Phys. C* **37** 117005
[12] Borland M *et al* 2014 *J. Synchrotron Rad.* **21** 912-936
[13] Jiao Y and Xu G 2015 *Chin. Phys. C* **39** 067004
[14] Farvacque L *et al* 2013 *Proc. IPAC'13* MOPEA008 (*Shanghai, China*)
[15] Borland M, Sajaev V, Sun Y and Xiao A 2015 *Proc. IPAC'15* TUPJE063 (*Richmond, USA*)
[16] Xu G, Jiao Y and Peng Y 2016 *Chin. Phys. C* **40** 027001
[17] Willeke F 2015 *Proc. IPAC'15* MOYGB3 (*Richmond, USA*)

[18] Terebilo A 2001 SLAC-PUB-8732

[19] Nadolski L and Laskar J 2003 *Phys. Rev. ST Accel. Beams* 6 114801

[20] Xu G *et al* 2016 *Proc. IPAC'16* WEOAA02 (*Busan, Korea*)

[21] Aiba M *et al* 2015 *Phys. Rev. ST Accel. Beams* **18** 020701

[22] Jiao Y 2016 *Chin. Phys. C* **40** 077002

[23] Bazarov I and Sinclair C K 2005 *Phys. Rev. ST Accel. Beams* **8** 034202

[24] Yang L *et al* 2009 *Nucl. Instrum. Methods Phys. Res.* A **609** 50

[25] Borland M *et al* 2009 *Proc. PAC'09* TH6PFP062 (*Vancouver, Canada*)

[26] Yang L, L Y, Guo W and Krinsky S 2011 *Phys. Rev. ST Accel. Beams* **14** 054001

[27] Gao W, Wang L and Li W 2011 *Phys. Rev. ST Accel. Beams* **14** 094001

[28] Bai Z, Wang L and L W 2011 *Proc. IPAC'11* WEPC109 (*San Sebastián, Spain*)

[29] Pang X and Rybarcyk L J 2014 *Nucl. Instrum. Methods Phys. Res.* A **741** 124

[30] Huang X and Safranek J 2014 *Nucl. Instrum. Methods Phys. Res.* A **757** 48

[31] Deb K *et al* 2002 *IEEE Transactions on Evolutionary Computation* **6** 182

[32] Laskar J 2003 *Proc. PAC'03* WOAB001 (*Portland, USA*)